# A Kac CROW Delay Line

M. Sumetsky

OFS Laboratories

19 Schoolhouse Rd, Somerset, NJ 08873

sumetski@ofsoptics.com

*Abstract*— A low-loss CROW delay line with a weak inter-resonator coupling determined by the Kac matrix is dispersionless and can be easily impedance-matched by adjusting the coupling to the input/output waveguide.

## I. Introduction

Optical delay lines, unless designed for other applications (e.g., dispersion compensation), are supposed to conserve the pulse shape, i.e., to be dispersionless. A regular optical fiber delay line is typically short enough to have a negligible dispersion. However, miniature slow light delay lines based on series of coupled microresonators [1,2] usually possess strong dispersion and have to be properly designed to eliminate it.

This paper is concerned with modeling of CROW (Coupled Resonator Optical Waveguide) delay lines [3]. It is common to design CROWs in the tight-binding approximation [4-6], which is valid if the width of transmission band is much smaller than the characteristic spacing between eigenfrequencies of an individual resonator. Based on this approximation, a CROW with all equally-spaced eigenfrequencies is introduced below. The remarkable property of this device is that it exhibits the dispersionless delay of light. It is named the Kac CROW after famous mathematician Mark Kac.

Based on the tight-binding approximation, it is found that a CROW can be impedance-matched to the input/output waveguide by tuning the coupling between the CROW and waveguide. Comparison of the performance of the impedance-matched Kac CROW and uniform CROW with the same number of resonators shows the benefits of a Kac CROW in larger delay time, smaller losses, and the absence of dispersion.

## II. The Kac Matrix And Dispersionless CROW

In 1854 James Sylvester published a paper in which he suggested that the $N \times N$ tridiagonal matrix **K** with elements $K_{n,n}=\lambda$, $K_{n+1,n}=n$, and $K_{n,n+1}=N-n$ had the determinant $\lambda(\lambda^2-2^2)(\lambda^2-4^2)\ldots(\lambda^2-(N-1)^2)$ for odd $N$ and $(\lambda^2-1^2)(\lambda^2-3^2)\ldots(\lambda^2-(N-1)^2)$ for even $N$, and, thus, had the equally-spaced eigenvalues $\lambda-N+1$, $\lambda-N+3$,…, $\lambda+N+3$,…, $\lambda+N-1$. Silvester's idea was proven in 1946 by Mark Kac. Therefore, the matrix **K** is often called the Kac matrix (see [7,8] for review). Here, to determine the dispersionless CROW, we employ the tridiagonal symmetrized Kac matrix **Q**, which has $Q_{n,n}=\lambda$ and $Q_{n+1,n}=Q_{n,n+1}=[n(N-n)]^{1/2}$. Matrix **Q** is similar to **K** and, thus, has the same equally-spaced eigenvalues.

Let us consider a CROW consisting of $N$ microresonators and operating in reflection (Fig. 1). It is assumed that all resonators, if uncoupled from each other, have the same complex resonance frequency $v_0+(i/2)\gamma_0$ with the internal loss $\gamma_0$. In the tight-binding approximation, the transmission amplitude of this device, $A(v)$, is expressed through coupling constants $\delta_n$ between resonators $n$ and $n+1$ and resonance width $\Gamma$, which determines coupling of resonator 1 to the input/output waveguide. We have $A(v)=1-i\Gamma(\mathbf{R}^{-1})_{1,1}$, where **R** is the symmetric tridiagonal matrix with $R_{n+1,n+1}=v-v_0-(i/2)\gamma_0$, $R_{n+1,n}=R_{n,n+1}=\delta_n$, for $n=1,2,\ldots N-1$, and $R_{1,1}=v-v_0-(i/2)(\gamma_0+\Gamma)$ [4]. The eigenvalues of **R** are the eigenfrequencies of CROW.

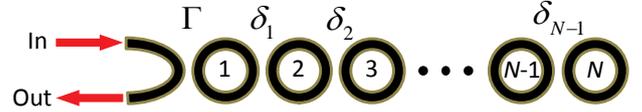

Fig. 1. Illustration of a CROW operating in reflection.

To arrive at a CROW, which is uncoupled from the waveguide ($\Gamma=0$) and has the equally-spaced eigenfrequencies, we can set $R_{n,n+1}=R_{n+1,n}=\Delta[n(N-n)]^{1/2}/2$, where $\Delta$ is the eigenfrequency spacing. However, the local bandwidth of such CROW strongly narrows near the edges. This shrinks the total transmission bandwidth and degrades the device performance. To solve the problem, we explore the symmetry of Kac matrix with respect to replacement $n \leftrightarrow N-n$ and utilize only the lower right-hand quadrant of this matrix by setting

$$\delta_n = \frac{\Delta}{4}\sqrt{N^2-n^2}, \quad n=1,2,\ldots,N-1. \quad (1)$$

In this case, the local bandwidth is maximized next to resonator 1 and monotonically decreases away from it. For small $\Gamma$, the eigenfrequencies of this matrix are $v_n=v_0+[(2n-N-1)+i\gamma_0]/2$ $n=1,2,\ldots N$, and the total bandwidth is $\Delta B=(N-1)\Delta$. Due to the equally-spaced eigenfrequencies, the defined CROW exhibits the dispersionless propagation of light as demonstrated below.

## III. Group delay, Transmission Power, And Pulse propagation

Generally, the group delay and transmission power spectra of a CROW with large number of resonators $N$ exhibit strong oscillations that corrupt the performance of this device. However, for a Kac CROW, oscillations vanish near the transmission band center if

$$\Gamma = 0.5\Delta B. \quad (2)$$

All the numerical results below are expressed in units of bandwidth $\Delta B$ (frequency) and inversed bandwidth $\Delta B^{-1}$ (time). For determinacy, it is assumed that the number of resonators $N=50$ and internal loss $\gamma_0=0.001\Delta B$. Fig. 2(a) and (b) show the group delay and transmission power spectra of a Kac CROW for $\Gamma=0.51\Delta B$ found from (2). The bold curves on the top of the spectra are the result of averaging over $0.03\Delta B$. In agreement with predictions of Section 1, the averaged group delay spectrum in the passband is close to a constant $N/\Delta B=50/\Delta B$.

The performance of the Kac CROW is investigated by considering the propagation of a Gaussian pulse having $0.25\Delta B$ spectral FWHM (Fig. 2(b)) and $1.76/\Delta B$ temporal FWHM (Fig. 2(c)). The results are shown in Fig. 2(c). The delay of the output pulse is equal to $49/\Delta B$ (28 bits), which is in a good agreement with the group delay found from Fig. 2(a). Importantly, the insert in Fig. 2(c) shows that the broadening of the output pulse is practically undetectable.

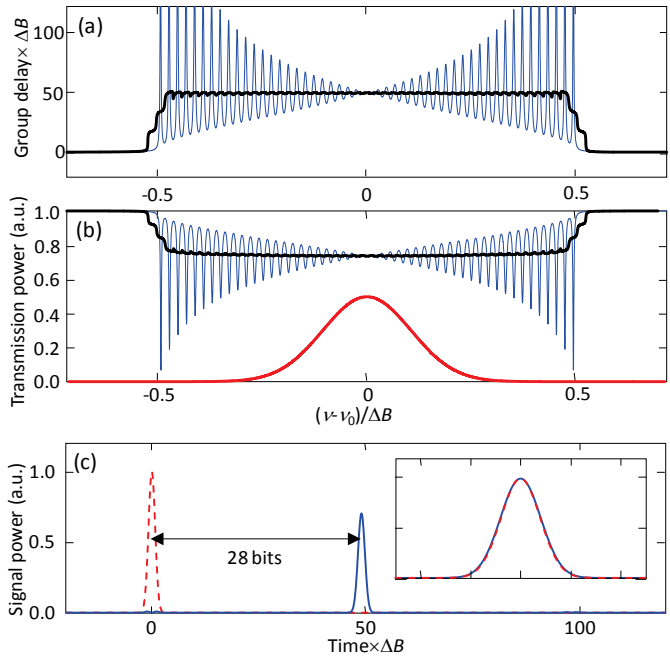

Fig. 2. (a) – Group delay and (b) – transmission power of a Kac CROW. Bold curve at the bottom of (b) is the power spectrum of the input Gaussian pulse. Bold curves at the top of spectra are the result of averaging over the $0.03\Delta B$ interval. (c) – Propagation of a Gaussian pulse through this device. Dashed curve – input signal, solid curve – output signal. Inset: comparison of the normalized input and output signals.

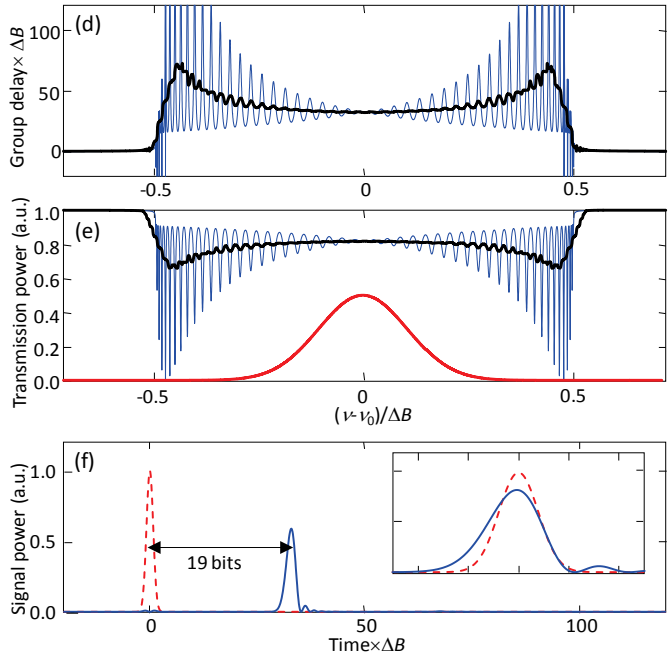

Fig. 3. (a) – Group delay and (b) – transmission power of a uniform CROW. Bold curve at the bottom of (b) is the power spectrum of the input Gaussian pulse. Bold curves at the top of spectra are the result of averaging over the $0.03\Delta B$ interval. (c) – Propagation of the Gaussian pulse through this device. Dashed curve – input signal, solid curve – output signal. Inset: comparison of the normalized input and output signals.

For comparison, Fig. 3 shows the results of similar calculations performed for a uniform CROW with $N$=50. In this case, the spectral oscillations in the middle of the transmission band vanish, again, under the condition of Eq. (2). The performance of this device is significantly worse than the performance of the Kac CROW shown in Fig. 2. The delay time $033/\Delta B$ (19 bits) is considerably smaller, while the attenuation of the pulse is greater and the pulse distortion shown in the inset of Fig. 3(c) is significant.

At the edges of transmission band in Fig. 3(a), the group delay experiences oscillations with negative values. These regions correspond to the under-coupled resonances which are characterized by the superluminal propagation and zero averaged group velocity. The spectral width of these regions grows with the internal loss $\gamma_0$. These regions do not show up for the Kac CROW due to the relatively small $\gamma_0$ corresponding to the over-coupled resonances along the entire transmission bandwidth. In a representative case of $\Delta B$=50 GHz (when the spectral FWHM of the Gaussian pulses in Fig. 2 and 3 is 12.5 GHz) we have $\gamma_0$=0.001$\Delta B$=50 MHz. At telecommunication wavelengths, this corresponds to the intrinsic Q-factor of resonators $\sim 10^6$. This Q-factor is achievable for planar microresonators [1,2] and is well within the range of the SNAP technology [9].

## IV. CONCLUSIONS

In this report, a Kac CROW delay line is introduced and investigated. It is shown that for relatively small internal losses this device is dispersionless within the entire transmission bandwidth. The inter-resonator coupling parameters and coupling to the input/output waveguide of a Kac CROW are determined by simple equations (1) and (2). Numerical modeling shows that the locally averaged group delay of a Kac CROW are close to a constant over the whole transmission bandwidth. As a result, a pulse, which has the spectral bandwidth comparable to the transmission bandwidth, experiences multi-bit delay without broadening. It is believed that the experimental realization of a Kac CROW is feasible with modern photonics fabrication technologies [1,2,9].